\documentclass[journal]{IEEEtran}
\usepackage{amsmath}
\usepackage{amssymb}
\usepackage{amsfonts}
\usepackage{graphicx}
\usepackage{epsfig}
\usepackage{epstopdf}
\graphicspath{{./figures/}}
\usepackage{subfigure}
\usepackage{bm}
\usepackage{multirow}
\usepackage{float}
\usepackage[linesnumbered, ruled]{algorithm2e}
\SetKwRepeat{Do}{do}{while}
\usepackage{booktabs}
\usepackage{nomencl}
\usepackage{cite}
\makenomenclature
\begin{document}
\title{Identify Critical Branches with Cascading Failure Chain Statistics and Hypertext-Induced Topic Search Algorithm}
\author{Chao~Luo,~\IEEEmembership{Student Member, IEEE,}
        Jun~Yang,~\IEEEmembership{Member, IEEE,}
        Jun~Yan,~\IEEEmembership{Student Member, IEEE,}
        Yufei~Tang,~\IEEEmembership{Member, IEEE,}
        Yuanzhang~Sun,~\IEEEmembership{Senior Member, IEEE}
        and Haibo~He,~\IEEEmembership{Senior Member, IEEE}
\thanks{Chao Luo, Jun Yang and Yuanzhang Sun are with the School of Electrical Engineering, Wuhan University, Wuhan, Hubei, 430072, China (E-mail: \{luochao\underline{ }2011, JYang\}@whu.edu.cn, yzs@tsinghua.edu.cn).}
\thanks{Jun Yan and Haibo He are with the Department of Electrical, Computer \& Biomedical Engineering, University of Rhode Island, Kingston, RI, 02881 USA (E-mail: \{jyan, he\}@ele.uri.edu).}
\thanks{Yufei Tang is with the Department of Computer \& Electrical Engineering and Computer Science, Florida Atlantic University, Boca Raton, FL 33431 USA (email: tangy@fau.edu).}
}
\maketitle

\begin{abstract}
An effective way to suppress the cascading failure risk is the branch capacity upgrade, whose optimal decision making, however, may incur high computational burden.
A practical way is to find out some critical branches as the candidates in advance.
This paper proposes a simulation data oriented approach to identify the critical branches with higher importance in cascading failure propagation.
First, a concept of cascading failure chain (CFC) is introduced and numerous samples of CFC are generated with an AC power flow based cascading failure simulator.
Then, a directed weighted graph is constructed, whose edges denotes the severities of branch interactions.
Third, the weighted hypertext-induced topic search (HITS) algorithm is used to rate and rank this graph's vertices, through which the critical branches can be identified accordingly.
Validations on IEEE 118-bus and RTS 96-bus systems show that the proposed approach can identify critical branches whose capacity upgrades suppress cascading failure risk more greatly.
Moreover, it is also shown that structural importance of a branch does not agree with its importance in cascading failure, which indicates the effectiveness of the proposed approach compared with structure vulnerabilities based identifying methods.

\end{abstract}

\begin{IEEEkeywords}
cascading failure; critical branches; interaction graph; HITS algorithm
\end{IEEEkeywords}
\IEEEpeerreviewmaketitle

\section{Introduction}
\IEEEPARstart{M}{odern} power grid is a complex infrastructure system with wide spatial span and huge amount of equipments.
Its security is threatened by many factors such as 1)various unexpected external disturbances, e.g., the fluctuations of load and renewable energy \cite{Wind_Power_Penetration_Robust_LFC}, the contingencies induced by equipment aging \cite{Failure_Rate_Model} or extreme weather \cite{Extreme_Weather_Cadini2017A}, and the recent cyber attacks \cite{Ukraine_Blackout2015}\cite{Jun_Yan_Attack_Review}; 2)some inherent mechanisms in power grid that may cause serious dependent outages, e.g., power flow transfer induced branch overloading.
Thus, power grid is prone to cascading failure and is faced with major blackout risk.
In fact, this has been verified by the massive historical records of blackouts in the world, e.g., the ones in United States and Canada on August 17th \cite{03AmericaBlackout2005}, in Europe on Dec. 4th, 2006 \cite{Europe2006Report}, in Brazil on Feb. 4th, 2011 \cite{09Brazil_Blackout} and in India on Jul. 30th \& 31th, 2012 \cite{India2012Report}.
Thus, modeling, understanding and mitigation of cascading failure have drew much attention recently \cite{Benchmarking_Validation2016}.
On the basis of constantly emerging of cascading failure models, e.g., the branching process models \cite{Janghoon_Kim2010BranchingProcessModel}\cite{Multi_Type_Branching}, the ORNL-PSerc-Alaska (OPA) models \cite{carreras2004complex,Shengwei_Mei2009improved_OPA,Junjian_Qi2013slow_process}, transient analysis based models \cite{Jiajia_Song2014dynamic_model}\cite{ChaoLuo_NAPS2016} and the multi-timescale models \cite{Pierre2015Two-Level}\cite{Rui_Yao2015Multi-Timescale}, researchers seek to develop effective countermeasures to alleviate the threats of cascading failure.

Since the cascaded branch overloading is widely accept as a major mechanism of cascading failure \cite{2005Blackout_Prevention_Overview}\cite{2006Anatomy_Blackout}, branch capacity upgrades (BCU) is viewed as an effective way to suppress the cascading failure risk (CFR) \cite{2017Dobson_Cascading}.
Several probabilistic BCU models have been proposed respectively in \cite{Karimi2015ProbabilisticTransmissionExpansion,Karimi2016,Shortle2014TEP,2015HighThroughputComputing} to optimize the upgrade plan (location and size) for the sake of minimizing the CFR or blackout probabilities.
Mathematically, the intractableness of this mixed integer simulation-optimization problem lies on 1)the time-consuming simulation in objective function evaluation and 2)the huge feasible region by mixed integer variables.
Although some heuristic methods have been utilized, e.g., the PSO algorithm \cite{Karimi2015ProbabilisticTransmissionExpansion}\cite{Karimi2016}, the tabu search \cite{Shortle2014TEP} and the high throughout computing platform based pattern search algorithm \cite{2015HighThroughputComputing},
they may still cause extremely high computational burdens when being applied in large power grid.
A practical way to address this issue is to reduce the searching space through choosing some critical branches as the candidates, which should have higher importance in cascading failure propagation and thereby whose upgrades can suppress the CFR more greatly.
Thus, how to search for these critical branches remains an issue worthing exploring.

The concept of critical branches are interpreted from various viewpoints.
For instance, within the application of complex network theory in power grid, it is assumed that the branches with higher structural importance are more critical.
Accordingly, some indices, e.g., vertices degree, edge betweenness \cite{2007Identification_of_Vulnerable_Lines}, electrical hybrid flow betweenness \cite{2015Hybrid_flow_betweenness}, and the extended betweenness \cite{Extended_Betweenness1}\cite{Extended_Betweenness2}, are introduced to rate and rank the branches.
However, these approaches emphasize more on the static structure vulnerabilities of power grid and ignore some basic characteristics, e.g., the Kirchhoff laws and the impact of supervision and control scheme of power grid.
Thus, the complete agreement between the structure importance of a branch and its importance in cascading failure can not be guaranteed.
Meanwhile, some other literatures define the critical branches as the ones whose outages will cause more serious violations in the remaining part of power grid.
Therefore, various searching approaches have been proposed in the on-line contingency analysis scheme such as the performance indices introduced by Ejebe and Wollenberg \cite{Automatic_Contingency_Selection1979}, the power transfer distribution factor (PTDF) based method \cite{Distribution_Factors}, the genetic algorithm based heuristic method \cite{Nims1997Contingency} and the electrical distance method \cite{Electrical_Distance2016}.
In fact, the branches found by these methods can only be viewed as the serious initial outages.
However, impacts of possible post-contingency cascading process are not considered.
Thus, the ones that play the exactly important roles in cascading failure propagation may be missed.

Essentially, cascading failure is a sequence of dependent outages caused by equipment interactions.
Thus, to make more correct and effective identification, the branch behaviors during cascading process should be considered reasonably.
Some researchers extract useful information directly from utilities' blackout records or from simulation results.
Ref. \cite{2012Vulnerabilities_Carreras} constructed a synchronization matrix from simulation results of OPA model to identify the branches with higher overloading probabilities.
Statistics of cascading branch outages spreading was analysed in \cite{dobson2015obtaining} using the historical utility records.
A line interaction graph was built in \cite{2013Paul_Hines_Dual_Graph} to provide useful insight into how lines interact with each other.
Further, a series of branch interaction model were proposed in \cite{2016Junjian_Qi_Interaction}\cite{2016Paul_Hines_Influence_Graph}, which can generate similar statistics of branch failures with the ones by original simulators.
Though, these works open a door to fully investigate the interaction between transmission branches during cascading failures, further discussions are still needed, e.g., how to integrally consider both the frequency and severity of branch interaction.
Inspired by the aforementioned achievements, we have made some works list as follows:
1)We use a standard structure of \emph{cascading failure chain} (CFC) to depict the cascading process.
Then, large amount of CFC samples are generated with an AC power flow based cascading failure simulator.
2)Based on the stochastic features of CFC, a directed weighted graph is constructed to quantify the interactive influences between branches with the consideration of both their occurrence frequency and severity.
3)The well known Hypertext-Induced Topic Search (HITS) algorithm is adopt to rate and rank the vertices of the graph, through which the critical branches of the original power grid can be identified equivalently;
4)The strategies of self-validation and cross-validation are adopt to validate the identification results of the proposed approach, which is shown to be able to identify the critical branches whose capacity upgrades can suppress CFR more greatly.
Moreover, it is also verified that structure vulnerabilities based importance of a branch does not agree with its importance in cascading failure propagation, which indicates the effectiveness of the proposed approach as well.

The rest of this paper is organised as follows:
Section II lists the conventional structure vulnerability based metrics as comparison;
Section III introduces the specifics of proposed identification approach and the validating method.
Case studies are presented in Section IV;
Section V draws the conclusion and gives discussions.
\section{Structure Vulnerability Based Metrics to Identify the Critical Branches}
Identifying the critical branches with metric from the perspective of structure vulnerability is a hot topic.
The metric \emph{betweenness}, which stems from the complex network theory, has been widely used to evaluate the structural importance of transmission branches in power grid.
Moreover, some modified metrics that combines the node pair concept of betweenness and electrical characteristics of power grids are proposed, e.g., the electrical betweenness and the extended betweenness \cite{2015Hybrid_flow_betweenness}\cite{Extended_Betweenness1}\cite{Extended_Betweenness2}.
Specifics of these metrics are introduced as follows.

\subsubsection{\textbf{Betweenness}}
Regardless of generators and loads, the power grid can be abstracted to be an undirected graph, whose vertices and edges represent the nodes and transmission branches of power grid respectively.
The betweenness of a edges is proposed as follows,
\begin{equation} \label{eq:betweenness}
B_1(l)=\sum_{i\in \bm{N},j\in \bm{N}}\frac{\sigma_{ij}(l)}{\sigma_{ij}} \quad , l\in \bm{B}
\end{equation}
where for any transmission branch (edge) $l$ that belongs to the branch set $\bm{B}$, $B_1(l)$ denotes its betweenness, and $\bm{N}$ stands for the electrical nodes (vertices) set.
For any node pair $i$-$j$, $\sigma_{ij}$ is the number of the shortest pathes between $i$ and $j$, and $\sigma_{ij}(l)$ is the number of the shortest pathes that go through branch $l$.

\subsubsection{\textbf{Electrical betweenness}}
The betweenness metric in (\ref{eq:betweenness}) is a pure topology based one, which does not consider the electrical characteristics of power grid, e.g., the power flow distribution and the interactive relationship between nodes connected with generators and load.
Thus, an electrical betweenness is proposed in   shown as follows.
\begin{equation}\label{eq:electrical_betweenness}
 B_2(l)=\sum_{i\in \bm{N}_G, j\in \bm{N}_D,i\neq j}\sqrt{W_iW_j}|I_{ij}(l)| \quad , l\in \bm{B}
\end{equation}
where $B_2(l)$ denotes the electrical betweenness metric of branch $l$, $\bm{N}_G$ is the set of nodes that are connected with generators, while $\bm{N}_D$ for the set of nodes connected with loads.
For any node pair $i$-$j$ that $i\in \bm{N}_G$ and $j\in \bm{N}_D$, $W_i$ denotes the maximum or real time output capacity of generators connected with node $i$, while $W_j$ denotes the maximum capacity of loads connected with node $j$.
$I_{ij}(l)$ is the current in branch $l$ when 1.0 p.u. current is injected into node $i$ meanwhile 1.0 p.u. current is withdrawn from node $j$.
This electrical betweenness metric considers the influence of generator and load capacity and power distribution characteristics, and hence seems to reflect the inherent feature of power grid with a more comprehensive way compared with the metric in (\ref{eq:betweenness}).

\subsubsection{\textbf{Extended betweenness}}
To combine the power flow distribution factors (PTDFs) with topological analysis, E.Bompard et al. proposed a novel \emph{extended betweenness} metric to analyze the structural stability of power grid, which can be calculated as follows.
Assume that each transmission branch has a designed limit $P_{max}(l)$, a pairwise power transmission capacity betweeen generation node $i$ and load node $j$ when the first branch in the grid reaches its limitation is defined as
\begin{equation}
P_{ij}=\min_{l\in \bm{B}}(\frac{P_{max}(l)}{|F_{i}(l)-F_{j}(l)|}), i\in \bm{N}_G, j\in \bm{N}_D
\end{equation}
where $F_i(l)$ and $F_j(l)$ are the power flow on branch $l$ when a unit power is injected into the generation node $i$ or the load node $j$ and withdrawn from the slack node, respectively.
Then, the extended betweenness for a branch $l$ is defined as the overall power transmitted across branch $l$ in a power grid, which is formulated as follows.
\begin{equation}
B_3(l) = \max \{T_P(l),T_N(l)\}
\end{equation}
where $T_P(l)=\sum_{i\in \bm{N}_G, j\in \bm{N}_D}\max\{F_{i}(l)-F_{j}(l),0\}P_{ij}$ and $T_N(l)=\sum_{i\in \bm{N}_G, j\in \bm{N}_D}|\min\{F_{i}(l)-F_{j}(l),0\}|P_{ij}$
are the positive directed and negative directed summated power flow on branch $l$ with respective to the transmission limit of each node pair $i$-$j$, i.e., $P_{ij}$, respectively.

All these three metrics are used to quantify the structural importance of branches.
It is believed by the complex network theory that the higher metric values indicate the higher important role in
sustaining the transmission efficiency of the topological structure.
Further comparisons between these methods with the proposed method is presented in Section IV.
\section{Cascading Failure Chain Statistics Based Critical Branches Identification}
The main task of this paper is to identify the critical branches with statistical information extracted from power grid behaviours in cascading failure process.
In what follows, specifics of the proposed approach are introduced.
\subsection{AC power flow based cascading failure simulation model}
Mechanisms of cascading failure are particularly complex, which makes it hard to model the cascading process with $100\%$ accuracy.
Without loss of generality, we consider two widely accept mechanisms, i.e., the cascading overloading and hidden failure of branches, and propose an AC power flow based simulation model, whose main components are presented here.
\subsubsection{\textbf{Selection of initial outages}}
We specify that the initial disturbances to trigger the cascading process are the branch outages, which can be selected in terms of their own independent failure probabilities \cite{Junjian_Qi2013slow_process} or can be any N-k contingencies, e.g., N-2 in \cite{Jiajia_Song2014dynamic_model}.
\subsubsection{\textbf{Selection of sequent outages}}
It is assumed here that branches fail due to either overloading or hidden failure.
In particular, when the load flow of branch $l$, denoted as $f_l$, is below its long term thermal limit $f_{lim1,l}$, its failure probability, denoted as $p_l$, is $0$.
If $f_l$ exceeds the short term emergent thermal limit, denoted as $f_{lim2,l}$, branch $l$ is considered to be in an emergent situation and should be cut off immediately to avoid the structural damages, thus, $p_l$ equals $1$ now.
Otherwise, when $f_l$ locates between $f_{lim1,l}$ and $f_{lim2,l}$, $p_l$ is formulated as a linear function of $f_i$ shown as follows.
\begin{equation}\label{eq:branch_failure_probability}
p_l=\begin{cases}
0\quad\quad\quad\quad\quad\quad  f_l\leq f_{lim1,l}\\
\frac{f_l-f_{lim1,l}}{f_{lim2,l}-f_{lim1,l}} \quad f_{lim1,l}<f_l\leq f_{lim2,l},\quad \forall l\in \bm{B}\\
1\quad\quad\quad\quad\quad\quad  f_l>f_{lim2,l}
\end{cases}
\end{equation}

Once occurring, operators should redispatch the system to eliminate the branch overloading.
However, the sagging caused tree flash and thermal effect caused aging failure may lead to the unexpected tripping of overloaded branches.
Thus, (\ref{eq:branch_failure_probability}) is a high-level probabilistic model to depict the uncertainties of the competition between undesigned thermal effect caused branch tripping and operators' actions \cite{Pierre_Probability_2015}.
Accordingly, in simulations, once obtaining the steady state of power system, we use (\ref{eq:branch_failure_probability}) to calculate the failure probabilities of branches, and then determine the sequent outages with random sampling.
Moreover, adjacent branches of the selected ones assumed to be exposed to hidden failures \cite{Chen2005cascading}.
The sequent outages by hidden failure are also randomly selected from the exposed branches in terms of their own hidden failure probability $p_{hi}$.
\subsubsection{\textbf{Topology updating}}
The topology information will be updated once some failed branches are removed.
Perform the connectivity checking for the whole power grid.
If the network breaks into several parts, the sub-grids are reserved, within which cascading failure simulations will still keep going on.
\subsubsection{\textbf{Power flow calculation}}
Post-contingency state is calculated by solving AC power flow model.
First, rebalance the active power deviation with appropriate generator ramping and load shedding (if needed).
Then, an AC power flow model with multi slack buses is proposed as follows.
\begin{subequations}
\begin{alignat}{2}
&\sum_{g\in \bm{G}_i}(P_{Gg}+r_g P_{loss})-\sum_{d\in \bm{L}_i}P_{Ld}-Re(\dot{V}_i\sum_{i\in \bm{N}}Y_{ij}^{*}\dot{V}_j^{*})=0  \\
&\sum_{g\in \bm{G}_i}Q_{Gg}-\sum_{d\in \bm{L}_i}Q_{Ld}-Im(\dot{V}_i\sum_{i\in \bm{N}}Y_{ij}^{*}\dot{V}_j^{*})=0
\end{alignat}
\end{subequations}
where $P_{Gg}$ and $Q_{Gg}$ are the active and reactive output of generator $g$ respectively; $P_{Ld}$ and $Q_{Ld}$ are
the the active and reactive power of load $d$; for each node $i$, $\dot{Y}_i$ is its voltage, $\bm{G}_i$ and $\bm{L}_i$ are sets of generators and loads connected with node $i$ respectively. $Y_{ij}$ the is mutual admittance between node $i$ and node $j$. Each generator is assigned an coefficient $r_g$ to share the transmission loss, denoted as $P_{loss}$.
Netwon-Raphson method is used to solve this problem.
Once the power flow model does not converge, it is viewed that voltage collapse occurs, then a load shedding strategy proposed in \cite{2016P_Rezaei_Thesis} is used to obtain a feasible power flow solution.
\subsubsection{\textbf{Emergent dispatch}}
The situation that no branch is selected in the random selection of sequent outages indicates that the operators succeed in pulling the power grid back into a secure state with no violations of the constraints of branch load flow and nodal voltage amplitude. This emergent dispatch is modeled as an optimal load shedding shown as follows.
\begin{subequations}
\begin{alignat}{2}
&\max \limits_{k_d,P_{Gg},Q_{Gg},V_i}\quad \sum_{d \in \bm{L}}k_dP_{Ld} \\
\mbox{s.t.}
&\sum_{g\in \bm{G}_i}P_{Gg}-\sum_{d\in \bm{L}_i}k_dP_{Ld}-Re(\dot{V}_i\sum_{i\in \bm{N}}Y_{ij}^{*}\dot{V}_j^{*})=0, \\
&\sum_{g\in \bm{G}_i}Q_{Gg}-\sum_{d\in \bm{L}_i}k_dQ_{Ld}-Im(\dot{V}_i\sum_{i\in \bm{N}}Y_{ij}^{*}\dot{V}_j^{*})=0, \\
&-f_{lim1,l}\leq f_l \leq f_{lim1,l}, \quad \quad\quad\; \forall l\in \bm{B}\\
&P_{Gg,min}\leq P_{Gg}\leq P_{Gg,max},\quad \quad \forall g\in \bm{G}\\
&Q_{Gg,min}\leq Q_{Gg}\leq Q_{Gg,max}, \ \ \ \ \ \forall g\in \bm{G}\\
&0\leq k_d \leq 1, \qquad \qquad \qquad \qquad \quad \forall d\in \bm{L}\\
&V_{min,i}\leq |\dot{V}_i| \leq V_{max,i}, \qquad \qquad \ \forall i\in \bm{N}
\end{alignat}
\end{subequations}
where $\bm{N}$, $\bm{B}$, $\bm{G}$ and $\bm{L}$ denote the sets of nodes, branches, generators and loads respectively;
$k_d$ is its shedding ratio that locates in $[0,1]$ for load $d$.
$V_{min,i}$ and $V_{max,i}$ are the lower \& upper boundaries for the amplitude of $\dot{V}_i$ and usually take the value of 0.9 and 1.1 respectively.
The implementation of emergent dispatch is the end of cascading failure, and the amount of load shedding can be viewed as the severity metric by the cascading failures.
\begin{algorithm}
\caption{Pseudocode of Cascading Failure Simulator}
\textbf{Initialize}: denote pre-contingency operation point as $Sys$

Define buffers $S1$ and $S2$ that $S1=\{Sys\}$,$S2=\varnothing$

\emph{Selection of initial outages} in $Sys$

\Repeat{$S1=\varnothing$}{
Assignment: $S2\leftarrow\varnothing$

\For{$\forall$ island $s_i\in S1$}{
\emph{Power flow calculation} for island $s_i$

\emph{Selection of sequent outages} for island $s_i$

\If{None sequent outage occurs in island $s_i$}{
\emph{Emergent dispatch} for island $s_i$
}

\emph{Topology updating} for island $s_i$

\If{island $s_i$ splits}{Denote newly generated islands as $g_1$,$g_2$,$\cdot\cdot\cdot$,$g_n$

$S2\leftarrow S2\bigcup{\{g_1}\}\bigcup{\{g_2}\}\cdot\cdot\cdot\bigcup{\{g_n}\}$
}
}
Assignment: $S1\leftarrow S2$
}

Obtain statistics of cascading failure chains and load loss
\end{algorithm}
\subsubsection{\textbf{Simulation process}}
The initial outages are randomly selected and then used to trigger the cascading failure simulation to obtain the statistics of load loss.
The whole process is represented by the pseudo code shown in Algorithm 1.
Several tools can be adopt to visualized the risk of cascading failure, e.g., the load loss distribution \cite{Jiajia_Song2014dynamic_model}, the VaR\&CVaR indices \cite{Shengwei_Mei2009improved_OPA} and the segmented risk histogram \cite{2016Paul_Hines_Influence_Graph}.

\subsection{Structure of cascading failure chain and evaluation of branch interactions}
Equipment failures may cause sequential outages due to some inherent mechanisms, e.g., power flow transfer induced branch overloading.
Thus, the propagation process of cascading failure can be viewed as a sequence of dependent outages that successively weakens or degrades the power grid \cite{2017Dobson_Cascading}.
Accordingly, in this paper the whole process of a complete cascading failure is abstracted as a \emph{cascading failure chain} (CFC), in which the massive outage events are grouped into stages with respect to their sequential orders.
\begin{figure}
\centering
\includegraphics[width=0.50\textwidth]{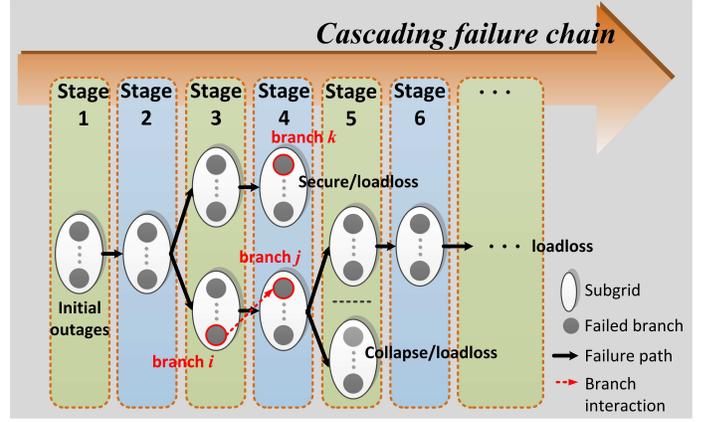}
\caption{Structure of cascading failure chain}\label{fig:CascadingFailureChain}
\end{figure}
Fig. \ref{fig:CascadingFailureChain} exhibits the typical structure of a CFC, which unfolds two points that should be noted when evaluating the branch interactions.
First, the power grid may be split into islands after some branch outages caused by overloading or rotor instability \cite{03AmericaBlackout2005}\cite{Europe2006Report}.
Thereafter, the cascading processes keep propagating within these islands respectively if appropriate emergent actions are not taken.
Thus, one should verify the cause-effect relationship between two branches in a CFC when evaluating their interactions.
For instance, in Fig. \ref{fig:CascadingFailureChain}, though branch $k$ fails immediately after branch $i$, there exists no cause-effect relationship between them since the island where branch $k$ fails does not derive from branch $i$ outage.
However, the similar works in \cite{2016Junjian_Qi_Interaction} and \cite{2016Paul_Hines_Influence_Graph} do not distinguish this difference and simply assume that any two branches have cause-effect relationship as far as they are in adjacent stages, which is not the reality.
Second, a specific dependent branch outage in CFC can be viewed as the trigger of subsequential cascading process, which may lead to different consequences finally, e.g., load shedding, returning to a secure state or collapsing completely.
Thus, for the sake of a more comprehensive evaluation, both the occurring frequency and the consequence severity of branch interactions should be considered.
However, aiming to construct the equivalent high level models in terms of outages statistics, the existing works \cite{2013Paul_Hines_Dual_Graph}\cite{2016Junjian_Qi_Interaction} \cite{2016Paul_Hines_Influence_Graph} only count the occurring frequency of branch interaction.
Moreover, it is noted that a CFC can be constructed through grouping the outages into stages, and
the outages data can be obtained from actual utility records \cite{dobson2015obtaining} or simulators.
In perticar, for the simulators with time marks, the outages can be grouped with respect to time closeness \cite{Impact_of_Topology2016TSG}\cite{2016Paul_Hines_Influence_Graph}\cite{1385363}.

As discussed above, in a CFC indexed with $s$, branch $l_i$ is viewed as the cause of branch $l_j$ only if branch $l_i$ fails in the previous stage of branch $l_j$ and meanwhile they have cause-effect relationship.
Thus, we denote this situation as $\sigma_{l_i,l_j}(s)=1$, and otherwise as $\sigma_{l_i,l_j}(s)=0$.
In this paper, the severity of interaction that branch $l_i$ outage causes branch $l_j$ outage in CFC $s$ is remarked as $M_{l_i\rightarrow l_j}(s)$ and evaluated as follows.
\begin{equation}\label{branch_impact}
M_{l_i\rightarrow l_j}(s)=
\begin{cases}
k_1\frac{e^{k_2\frac{Loss_{l_j}(s)}{L_T}}}{N_{l_i}(s)N_{l_j}(s)},&\mbox{$\sigma_{l_i,l_j}(s)=1$}\\
0,&\mbox{\, $\sigma_{l_i,l_j}(s)=0$}
\end{cases}
\end{equation}

When $\sigma_{l_i,l_j}(s)=0$, branch $l_i$ outage is not the cause of branch $l_j$ outage, thus $M_{l_i\rightarrow l_j}(s)$ takes the value of zero, while when $\sigma_{l_i,l_j}(s)=1$, an exponential utility function based metric is proposed here in terms of $Loss_{l_j}(s)$, denoting the total load loss along the whole subsequential process of branch $l_j$ outage.
$L_T$ is the total supplied load in the original power grid.
$k_1$ and $k_2$ are the non-negative scaling parameters.
$N_{l_i}(s)$ is the total number of branches that fail along with branch $l_i$ within the same island and in the same stage (including branch $l_i$).
It is the same with $N_{l_j}$.
It is obvious that under given $k_1$ and $k_2$, $M_{l_i\rightarrow l_j}(s)$ is a non-decreasing function of $Loss_{l_j}(s)$.
When $k_2=0$, $M_{l_i\rightarrow l_j}(s)$ can be viewed as an occurrence indicator of the event $\sigma_{l_i,l_j}(s)=1$, and thus reflects the occurring frequency in essences,
while when $k_2>0$, both the load loss severity and the occurrence frequency can be considered.
Besides, since there may be multi outages within the same island in a stage, $N_{l_i}(s)$ and $N_{l_j}(s)$ are introduced to evenly attribute the severity to all involved branches.

Further, if the set of all potential CFSs is denoted as $\Omega_{P}$, the total impacts of branch $l_i$ on branch $l_j$, denoted as $E_{l_i\rightarrow l_j}$, can be calculated as follows.
\begin{equation}\label{1}
E_{l_i\rightarrow l_j}=\sum_{s\in \Omega_{P}}prob_{l_i\rightarrow l_j}(s)M_{l_i\rightarrow l_j}(s)
\end{equation}
where $prob_{l_i\rightarrow l_j}(s)$ is the occurring probability of the event $\sigma_{l_i,l_j}(s)=1$ and can be theoretically calculated as $prob_{l_i\rightarrow l_j}(s)=p_{j}(s)\prod_{k}p_k(s)$ where $p_k(s)$ is the conditional probability of outages in previous stages of branch $l_j$ outage and $p_{j}(s)$ is failure probability of branch $l_j$ in CFC $s$.
However, since $\Omega_{P}$ is hard to be exhaustively enumerated, $E_{l_i\rightarrow l_j}$ is estimated through CFC sampling in this paper as follows.
\begin{equation}\label{1}
\hat{E}_{l_i\rightarrow l_j}=\frac{1}{N_s}\sum_{s=1}^{N_s}M_{l_i\rightarrow l_j}(s)
\end{equation}
where $N_s$ is the total number of CFC samples that are obtained with the cascading failure simulator proposed in this paper.

Finally, a directive weighted graph $\mathcal{G}$ can be constructed to represent the overall interactions between branches.
Vertices of $\mathcal{G}$ denote the branches in original power grid, and the directed edges of $\mathcal{G}$ denote the interactive relationship between branches.
Accordingly, $\mathcal{G}$ can be represented by the weighted adjacency  matrix $W$, whose element $w_{ij}$ can be determined as follow.
\begin{equation}\label{branch_impact}
w_{ij}=
\begin{cases}
\hat{E}_{l_i\rightarrow l_j},&i\neq j\\
0,&i=j
\end{cases}
\end{equation}

\subsection{HITS Algorithm}
\begin{figure}
  \centering
  \includegraphics[width=0.37\textwidth]{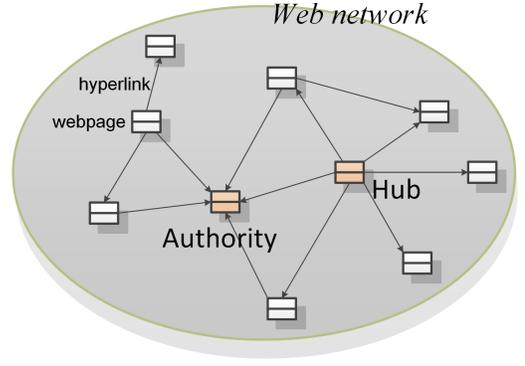}
  \caption{Illustration of authority and hub webs}\label{fig:HITS}
\end{figure}

In this paper, the well-known Hyper-Induced Topic Search (HITS) algorithm is adopt to rate and rank the vertices of $\mathcal{G}$, through which the critical branches can be identified accordingly.
HITS algorithm is a link analysis algorithm that was originally developed in \cite{Kleinberg1999Hubs} for a search engine to select the highly relevant web pages for a particular query.
As shown in Fig. \ref{fig:HITS}, the web network can be represented as a directed graph, whose vertices and edges denote the web pages and the hyperlink relationships between pages respectively.
The main concepts of HITS algorithm are the the hub, representing a page that points to many other pages and thus provides more accesses to the useful pages, and the authority, representing a page that was pointed by many other hubs and thus has more information relevant with the query theme.
Each node (web page) $i$ is assigned with two attributes, the \emph{hub value}, $hub_i$ and the \emph{authority value}, $auth_i$.
It is assumed that the authority value of node $i$ is reinforced by the hub values of nodes that point to $i$, and at the same time the hub value of node $i$ is reinforced by the authority values of the nodes pointed by $i$.
Thus, for the sake of calculating $hub_i$ and $auth_i$, iterations can be performed as $auth^{(k+1)}_i=\sum_{j:j\rightarrow i}hub_j^{(k)}$, $hub^{(k+1)}_i=\sum_{j:i\rightarrow j}auth^{(k+1)}_j$, where $j:j\Rightarrow i$ denotes the set of nodes that point to $i$, and the superscript $k$ denotes the iteration number \cite{Kleinberg1999Hubs}\cite{Kleinberg1999Authoritative}.

However, the aforementioned algorithm is a purely topological structure oriented one, and does not consider the edge weights of the network, which may denote the quality of web linkages, e.g., the occasional failing and the user preference of the linkages.
Therefore, a weighted HITS algorithm was introduced and theoretically analysed in \cite{HITS_Theoretical_Study}.
We utilize this revised formulation of HITS algorithm to analyse the branch interaction graph $\mathcal{G}$ built in the last subsection.
Specifically, each vertex $i$ (i.e., branch $i$) of $\mathcal{G}$ is assigned with two attributes, the impact severities of causing other branches' failing and being caused by other branches, which can be viewed as analogies to the concepts of $authority$ and $hub$ and thus are still denoted as $hub_i$ and $auth_i$ respectively.
Combined with the weighted adjacency matrix $W$ of $\mathcal{G}$, the updating strategies for $hub_i$ and $auth_i$ are shown as follows.
\begin{equation}
 auth^{(k+1)}_i=\sum_{j:j\rightarrow i}\frac{w_{ji}}{\sum_{p:j\rightarrow p}w_{jp}}hub_j^{(k)}
\end{equation}
\begin{equation}
 hub^{(k+1)}_i=\sum_{j:i\rightarrow j}\frac{w_{ij}}{\sum_{p:p\rightarrow j}w_{pj}}auth^{(k+1)}_j
\end{equation}
where $hub_j$ is attributed to the vertices pointed by vertex $j$ with the proportion of edge weights between them,
meanwhile $auth_j$ is also attributed to the vertices pointing to vertex $j$ with the same strategy.
Since $auth^{k}_i$ and $hub^{k}_i$ may become too large in the iteration process, a normalization for them is imposed at each step.
Details of the calculation process are shown in Algorithm 2.
In addition, the convergence of iteration is guaranteed under the condition that $\mathcal{G}$ is strongly connected \cite{HITS_Theoretical_Study}, thus the zero entries of $W$ can be replaced with a very small positive real number, which has little influence on the final results.
Finally, a composite metric can be formulated as follows
\begin{equation}\label{1}
K_i=\frac{1}{2}(auth_i+hub_i), \forall i \in \bm{B}
\end{equation}
where $K_i$ is the importance metric in cascading failure for branch $i$, which can comprehensively take into account the two attributes' effects.
The branches with larger $K_i$ values are viewed as having higher importance in propagation of cascading failures, thus the critical branches can be identified from the descending order of $K_i$ values.
\begin{algorithm}
\caption{Pseudocode of weighted HITS algorithm}
\SetKwInOut{Input}{Input}
\SetKwInOut{Output}{Output}
\Input  {$N\times N$ weighted adjacency matrix $W$ of the branch interaction graph $\mathcal{G}$}
\Output {the importance metric for each branch $K_i$}
Define $N\times 1$ vectors $A^{(k)}$, $H^{(k)}$, $k=0,1,2,\cdots$;

Entries of $A^{(k)}$ and $H^{(k)}$ are denoted as $auth_i^{(k)}$ and $hub_i^{(k)}$ respectively, $1\leq i\leq N$;

\textbf{Initialization}: $auth_i^{(0)}=1$, $hub_i^{(0)}=1$, $k=1$

\Repeat{$\|A^{(k)}-A^{(k-1)}\|_{\infty}+\|H^{(k)}-H^{(k-1)}\|_{\infty}<\varepsilon$}{

\For{$1\leq i\leq N$}{
$auth^{(k)}_i=\sum_{j:j\rightarrow i}\frac{w_{ji}}{\sum_{p:j\rightarrow p}w_{jp}}hub_j^{(k-1)}$

$hub^{(k)}_i=\sum_{j:i\rightarrow j}\frac{w_{ij}}{\sum_{p:p\rightarrow j}w_{pj}}auth^{(k)}_j$
}

normalize $A^{(k+1)}$ that $\|A^{(k+1)}\|_2=1$;

normalize $H^{(k+1)}$ that $\|H^{(k+1)}\|_2=1$;

Updating $k$: $k=k+1$
}

Assign: $K_i=\frac{1}{2}(auth_i^{(k)}+hub_i^{(k)}),1\leq i \leq N$

\end{algorithm}

\subsection{Validation for Identification Results}
Due to lacking rigorous analytical models, the identification results should be validated.
A simple but effective branch capacity expansion model has been used in the vulnerability analysis of complex network \cite{Wenli_Fan2016}\cite{JunYan_Self-Organizing_Map}, in which the capacity of a node/branch is set as $C_i=\alpha L_{i,0}$ where $L_{i,0}$ is its initial load and $\alpha$ is the system-wide tolerance coefficient bigger than $1.0$.
An external attack scheme is viewed more vulnerable if its severity decreases less when $\alpha$ increases.
Similarly, in this paper we formulate a simple branch capacity upgrade model as follows
\begin{subequations}\label{eq:branch_capacity_upgrade}
\begin{alignat}{2}
f_{lim1,l} = f_{lim1,l}+ \Delta C,\qquad l\in \bm{B}_C\\
f_{lim2,l} = f_{lim2,l}+ \Delta C ,\qquad l\in \bm{B}_C
\end{alignat}
\end{subequations}
where $\Delta C$ is the increment of branch capacity, and $\bm{B}_C$ denotes the set of branches to be upgraded.
It is assumed in this paper that when $\bm{B}_C$ is composed of more critical branches, the CFR can be suppressed more greatly.
Then, two strategies are adopt to perform the validations \cite{Benchmarking_Validation2016} as follows:
\subsubsection{Self-validation}
Based on the identification results of the proposed approach itself, comparisons are made to validate whether upgrading more critical branches will suppress cascading failure risk more greatly, and so as to validate the reasonability of the proposed metric $K_i$ to reflect the branch importance.
\subsubsection{Cross-validation}
Compare identification results of different approaches, e.g., the proposed one and the structure vulnerabilities based ones, and validate whether upgrading branches identified by the proposed approach seems more effective.
\section{Case Studies and Results}
\subsection{Tests on IEEE 118 bus system}
\begin{figure}
\centering
\includegraphics[width=0.4\textwidth]{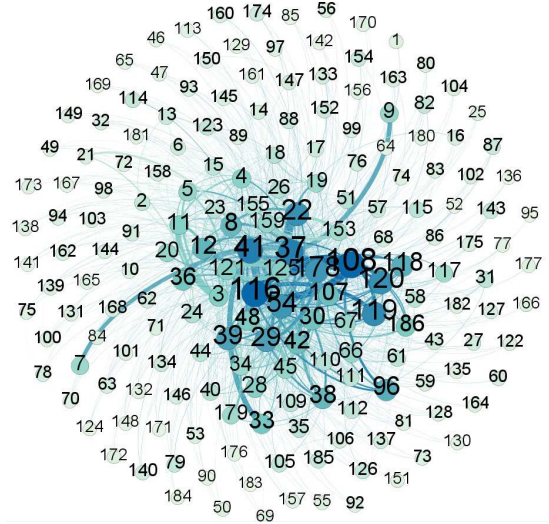}
\caption{Gephi based visualization of the graph $\mathcal{G}$ of IEEE 118 bus system. The size of a vertex label is proportional to its $K_i$ value (importance of branch), and the width of a edge is proportional to its weight value $w_{ij}$ (i.e., interaction severity between branches).}\label{fig:Gephi_graph}
\end{figure}
\begin{figure}
\centering
\subfigure[]{\includegraphics[width=0.5\textwidth]{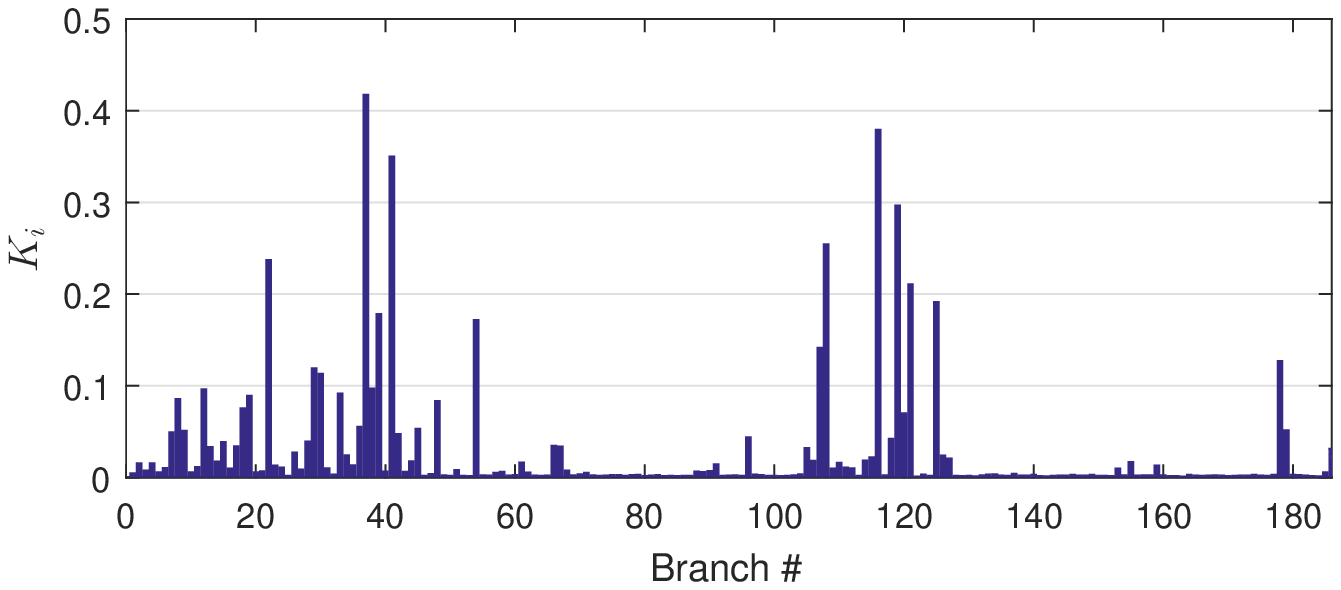}}
\subfigure[]{\includegraphics[width=0.5\textwidth]{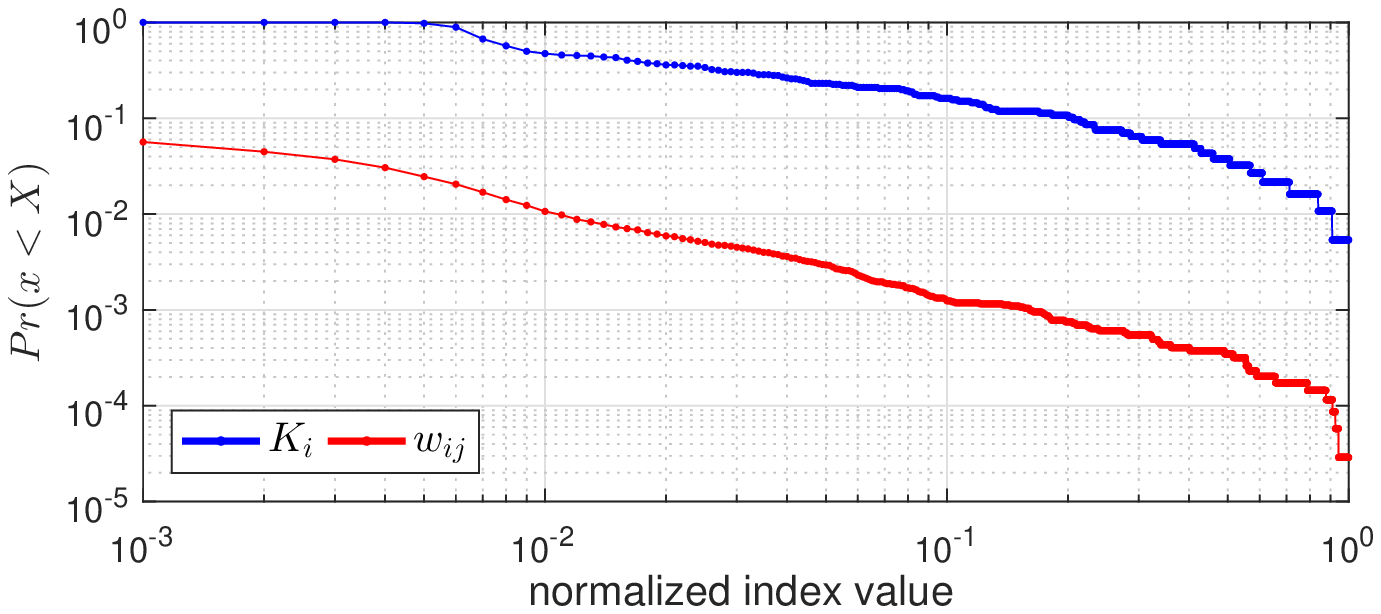}}
\caption{Identification results for IEEE 118 bus system, in which (a) shows the histogram of $K_i$ values across all branches marked from 1 to 186; (b) shows the distributions of $K_i$ and $w_{ij}$ which are normalized with min-max scaling.}\label{fig:index_distribution_IEEE118}
\end{figure}
Case studies are performed on the IEEE 118 bus system, whose data can be accessed from \cite{IEEE118_benchmark}.
It has $186$ branches and $3733$ MW load.
To increase the operation stress, each load is increased by $60\%$.
The values of $f_{lim1,l}$ are set to be 140 MW. and 450 MW for transmission lines and transformers respectively.
$V_{min,i}$ and $V_{max,i}$ take the value of 0.9 and 1.1 respectively for all nodes.
The hidden failure probabilities of branches are all set as 0.01.
The initial operation point is determined by optimal power flow model.
Initial outages are randomly selected from N-2 branch contingencies \cite{Jiajia_Song2014dynamic_model}, then numerous CFC samples are obtained through simulations.
The iteration number for each simulation is set as $1.0\times 10^5$,
Thereafter, critical branches are identified with the proposed approach, in which the parameters $k_1$ and $k_2$ are set as 6 and 3 respectively. The threshold $\varepsilon$ for the weighted HITS algorithm is $10^{-5}$.
In result validations, $\Delta C$ takes the value of 300MW.

Visualization of the graph $\mathcal{G}$ is implemented as shown in Fig. \ref{fig:Gephi_graph}, in which the size of vertex label is proportional to the corresponding branch's $K_i$ value, and the edge width is proportional to $w_{ij}$ value of the corresponding branch pair.
It is observed that the severities of most branch interactions are relatively low, while the strong interactions mainly exist between only a small part of branches, including branch 37, 41, 119, 108, 116, 121, etc..
Thus, it can be inferred that these branches are the critical hubs or authorities in the propagation process of cascading failure.
Further, the histogram of $K_i$ values of all 186 branches are shown in Fig. \ref{fig:index_distribution_IEEE118}(a).
It can be observed that only one branch has the $K_i$ value that is bigger than $0.4$, and there exist only 13 branches whose $K_i$ values exceed 0.1.
Apart from these high ranking branches, $K_i$ values of all other branches are below 0.1, moreover even nearly half of the branches are with $K_i$ values below 0.01.
For a more clear revelation, the distributions of normalized $K_i$ and $w_{ij}$ values are depicted in Fig. \ref{fig:index_distribution_IEEE118}(b).
It shows that both of these two metrics appear to have the characteristics of power law distribution, which, in other words, support the awareness that only a small part of the branches have much higher importance than others.
Thus, based on these observations, it can be verified that some branches may be more prone to be affected by other outages or the failures of themselves are more apt to cause sever sequential outages.
These branches play more important roles out of others in promoting the propagation of cascading failures, which can be identified effectively by the proposed approach.
\begin{figure}
\centering
\subfigure[]{\includegraphics[width=0.5\textwidth]{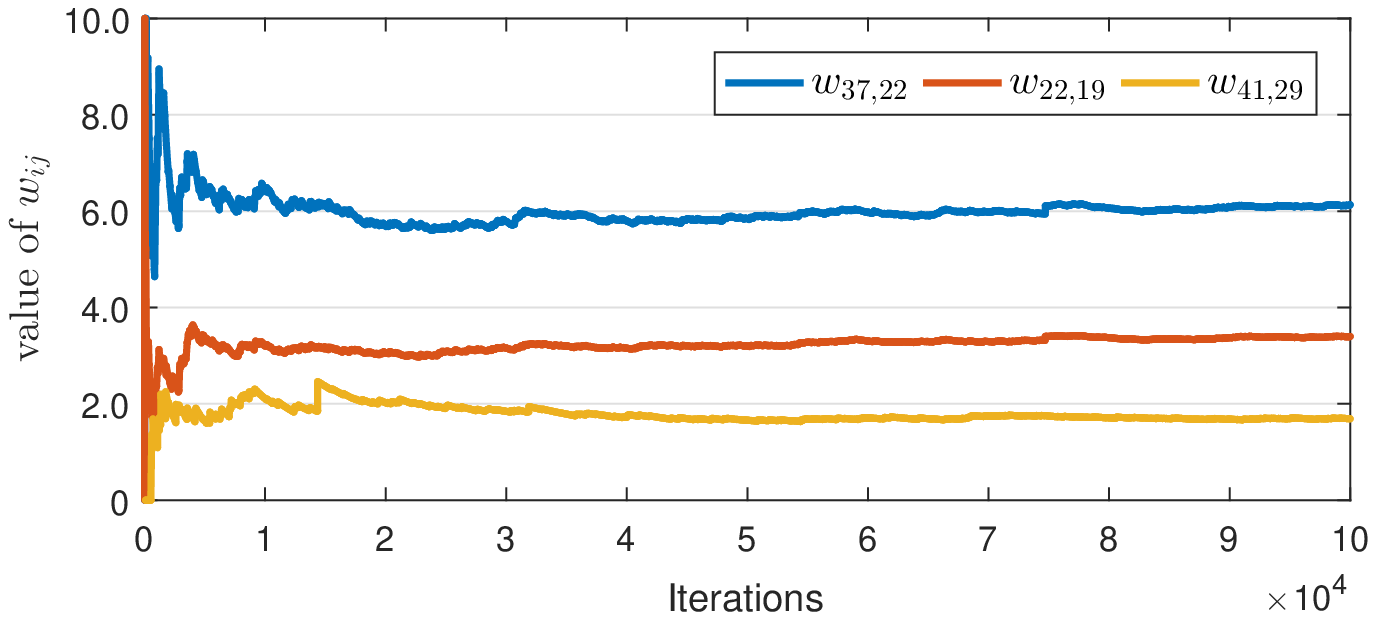}}
\subfigure[]{\includegraphics[width=0.5\textwidth]{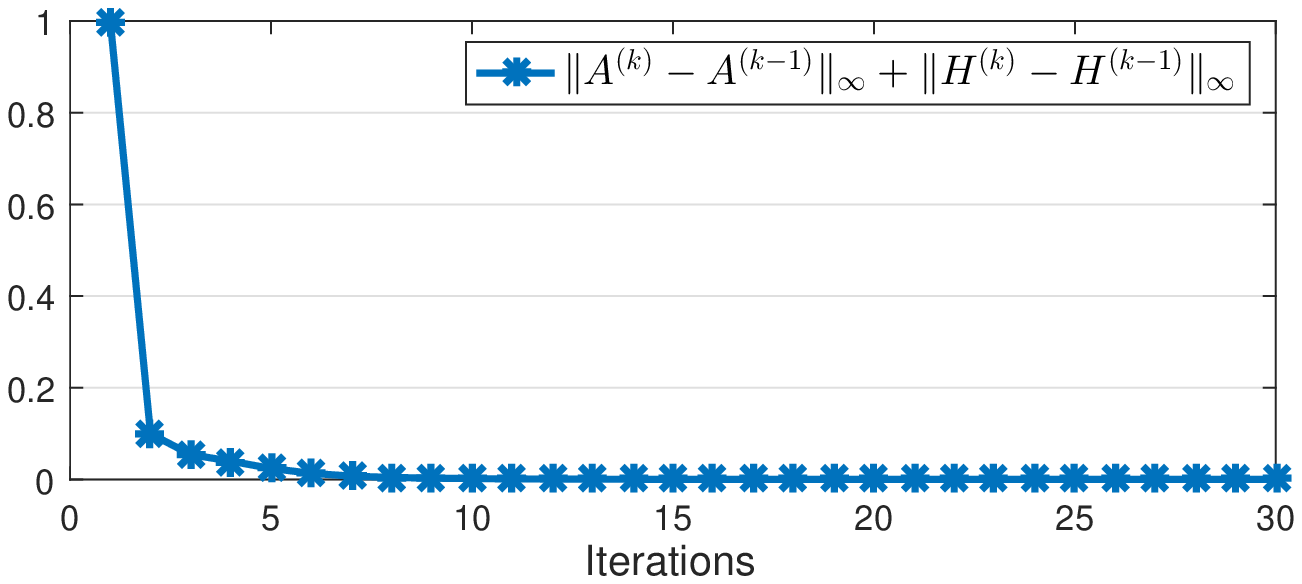}}
\caption{The converging characteristics of the proposed approach on IEEE 118 bus system. (a) shows the converging curves of some entries of $W$; (b) shows the converging curve of weighted HITS algorithm.}\label{fig:converging curves}
\end{figure}
In addition, Fig. \ref{fig:converging curves} shows the convergence characteristics of the proposed approach.
The converging curves of $w_{ij}$, taking $w_{37,22}$, $w_{22,19}$ and $w_{41,29}$ as examples, demonstrate the reliable performances of both cascading failure simulator and the proposed metric $w_{ij}$.
Meanwhile, as depicted in Fig. \ref{fig:converging curves}(b), the iterative error of the weighted HITS algorithm decreases below the threshold $\varepsilon$ after about 30 iterations, indicating the credibility of the proposed approach.
\begin{table}
\centering
\renewcommand{\arraystretch}{1.2}
\caption{Critical branch identification results by the proposed approach for IEEE 118-bus system}
\label{tab:Branch_Ranking_HITS_IEEE118}
\begin{tabular}{ccc|ccc}
\hline
\hline
Rank&Branch&$K_i$&Rank&Branch&$K_i$\\
\hline
1&37(8-30)&$0.4183$&21&18(13-15)&$0.0765$\\
2&116(69-75)&$0.3803$&22&120(75-77)&$0.0710$\\
3&41(23-32)&$0.3521$&23&36(30-17)&$0.0561$\\
4&119(69-77)&$0.2976$&24&45(19-34)&$0.0540$\\
5&108(69-70)&$0.2552$&25&179(32-113)&$0.0524$\\
6&22(16-17)&$0.2382$&26&9(9-10)&$0.0518$\\
7&121(77-78)&$0.2118$&...&...&...\\
8&125(79-80)&$0.1922$&175&129(82-83)&$2.57\times10^{-3}$\\
9&39(17-31)&$0.1793$&176&141(89-92)&$2.56\times10^{-3}$\\
10&54(30-38)&$0.1727$&177&83(51-58) &$2.52\times10^{-3}$\\
11&107(68-69)&$0.1425$&178&80(56-57) &$2.50\times10^{-3}$\\
12&178(17-113)&$0.1280$&179&101(62-67)&$2.49\times10^{-3}$\\
...&...&...&180&161(92-102)&$2.45\times10^{-3}$\\
15&38(26-30)&$0.0979$&181&170(105-107)&$2.42\times10^{-3}$\\
16&12(11-12)&$0.0973$&182&131(83-85)&$2.33\times10^{-3}$\\
17&33(25-27)&$0.0925$&183&142(89-92)&$2.28\times10^{-3}$\\
18&19(14-15)&$0.0899$&184&184(12-117)&$2.17\times10^{-3}$\\
19&8(8-5)&$0.0865$&185&122(78-79)&$2.05\times10^{-3}$\\
20&48(33-37)&$0.0842$&186&163(100-103)&$1.92\times10^{-3}$\\
\hline
\hline
\end{tabular}
\vspace{-3mm}
\end{table}
\begin{table}
\centering
\renewcommand{\arraystretch}{1.2}
\caption{Critical branch identification results by structure vulnerability based metrics for IEEE 118-bus system}\label{tab:Betweenness_Branch Ranking_IEEE118}
\begin{tabular}{cccccc}
\hline
\hline
\multicolumn{2}{c}{\emph{Betweenness}}&\multicolumn{2}{c}{\emph{Electrical Betweenness}}&\multicolumn{2}{c}{\emph{Extended Betweenness}}\\
\cmidrule(lr){1-2}\cmidrule(lr){3-4}\cmidrule(lr){5-6}
Rank&Branch&Rank&Branch&Rank&Branch\\
\hline
1&104(65-68)&1&104(65-68)&1&104(65-68)\\
2&96(38-65)&2&127(81-80)&2&126(68-81)\\
3&54(30-38)&3&126(68-81)&3&127(81-80)\\
4&126(68-81)&4&96(38-65)&4&96(38-65)\\
5&127(81-80)&5&54(30-38)&5&54(30-38)\\
6&37(8-30)&6&97(64-65)&6&30(23-24)\\
7&128(77-82)&7&102(65-66)&7&102(65-66)\\
8&36(30-17)&8&30(23-24)&8&128(77-82)\\
9&102(65-66)&9&128(77-82)&9&97(64-65)\\
10&8(8-5)&10&37(8-30)&10&8(8-5)\\
11&152(80-98)&11&119(69-77)&11&155(84-100)\\
12&97(64-65)&12&107(68-69)&12&163(100-103)\\
...&...&...&...&...&...\\
\hline
\hline
\end{tabular}
\vspace{-3mm}
\end{table}

To validate the identification results, several comparative case studies are made here.
First, whole 186 branches of IEEE 118-bus system are ranked in a descending order in terms of their $K_i$ values, from which three typical groups of branches are selected and shown in Tab. \ref{tab:Branch_Ranking_HITS_IEEE118}, i.e., \emph{top ranking branches}, which rank $1^{st}$ to $12^{th}$, \emph{middle ranking branches}, which rank $15^{th}$ to $26^{th}$, and \emph{bottom ranking branches}, which rank $175^{th}$ to $186^{th}$.
As listed in Tab. \ref{tab:Branch_Ranking_HITS_IEEE118}, $K_i$ values of these three groups of branches locate in the ranges $[0.1280,0.4183]$, $[0.0518, 0.0979]$ and $[1.92\times 10^{-3}, 2.57\times 10^{-3}]$ respectively, which are three typical different importance levels.
Meanwhile, as comparison, top 12 branches of the rankings in terms of betweenness, electrical betweenness and extended betweenness are respectively listed in Tab. \ref{tab:Betweenness_Branch Ranking_IEEE118}.
Then, based on the branch capacity upgrade model in (\ref{eq:branch_capacity_upgrade}), the self-validation and cross-validation for the identification results are made as follows.
\subsubsection{self-validation}
\begin{figure}
\centering
\subfigure[]{\includegraphics[width=0.5\textwidth]{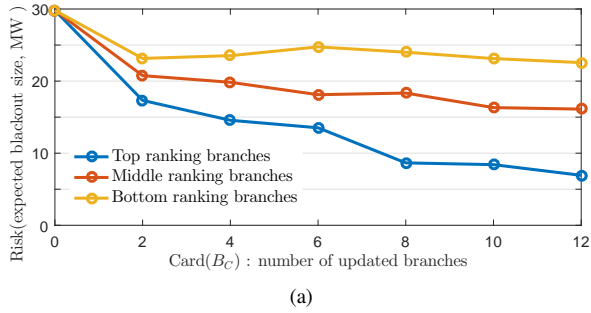}}
\subfigure[]{\includegraphics[width=0.5\textwidth]{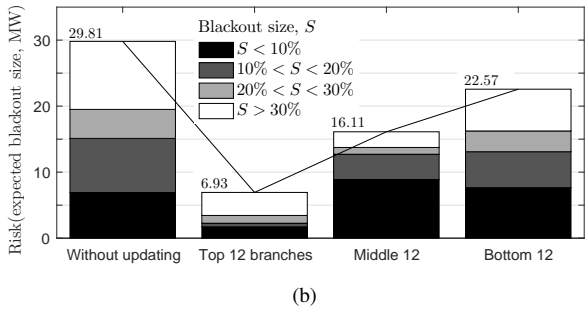}}
\caption{Results of self-validation on IEEE 118 bus system. (a) shows the performance comparisons of branch upgrade plans with top-ranking, middle-ranking and bottom-ranking branches identified by the proposed approach respectively; (b) shows the interval risk histograms of different upgrading plans when Card$(\bm{B}_C)=12$.}\label{fig:self_validation_IEEE118}
\end{figure}
We compare the CFR suppression performances when branches to be upgraded are selected from aforementioned three groups respectively.
Specifically, given the number of branches to be upgraded, which equals Card($\bm{B}_C$), top Card($\bm{B}_C$) branches of each branch group are upgraded respectively according to (\ref{eq:branch_capacity_upgrade}).
Then simulations are performed to calculate the CFR (estimated by the average expected load loss).
Fig. \ref{fig:self_validation_IEEE118}(a) shows the changes of CFR for all three branch groups when Card($\bm{B}_C$) increases from $0$ to $12$ with a step of 2.
CFR of original system is $29.81$ MW (Card($\bm{B}_C$)=0), and keeps reducing dramatically when more branches are upgraded (Card($\bm{B}_C$)$>$0) no matter which group the upgraded branches are selected from.
This indicates that branch capacity upgrade is an effective countermeasure to suppress CFR.
However, its performance varies when branches with different importance are upgraded.
It is observed from Fig. \ref{fig:self_validation_IEEE118}(a) that CFR decreases more greatly when top ranking branches are selected than other two groups of branches, meanwhile those bottom ranking branches have the lowest performance.
This observation corresponds to the ranking of these branches' $K_i$ values that upgrading branches with higher $K_i$ values seems more effective.
Thus, it can be verified that the proposed metric $K_i$ is reasonable to reflect the branches' importance in cascading failure propagation.
In addition, to further reveal the effectiveness of upgrading critical branches, Fig. \ref{fig:self_validation_IEEE118}(b) presents the interval risk histograms of a specific case that Card($\bm{B}_C$)=12.
It is shown that CFR can be reduced from 29.81 MW to 6.93 MW by upgrading top 12 branches, to 16.11 MW by upgrading middle 12 branches and to 22.57 MW by upgrading bottom 12 branches respectively.
Moreover, when top 12 branches are upgraded, risks of all intervals, no matter small blackouts (load loss$<10\%$), medium blackouts ($10\%<$load loss$<30\%$) or large blackouts (load loss$>30\%$), decreases significantly.
On the contrary, the risk of small blackouts does not show any reduction when middle or bottom 12 branches are selected.

\subsubsection{cross-validation}
\begin{figure}
\centering
\subfigure[]{\includegraphics[width=0.5\textwidth]{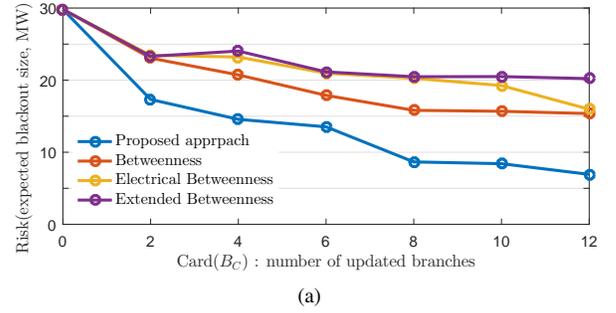}}
\subfigure[]{\includegraphics[width=0.5\textwidth]{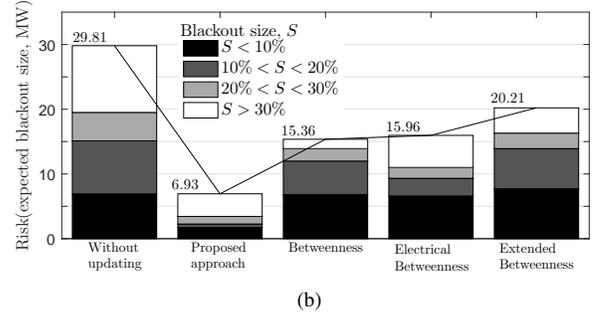}}
\caption{Results of cross-validation on IEEE 118 bus system. (a) shows the performance comparisons of uprading plans with top-ranking branches identified by the proposed approach and structure vulnerability based metrics respectively; (b) shows the interval risk histograms of different upgrading plans when Card$(\bm{B}_C)=12$.}\label{fig:cross_validation_IEEE118}
\end{figure}
We compare the CFR suppression performances when branches to be upgraded are selected from identification results of the proposed approach, betweenness, electrical betweenness and extended betweenness respectively.
Specifically, given the number of branches to be upgraded, Card($\bm{B}_C$), top Card($\bm{B}_C$) branches which are ranked by different approaches are upgraded respectively.
Fig. \ref{fig:cross_validation_IEEE118}(a) shows the changes of CFR for all identification approaches when Card($\bm{B}_C$) increases from 0 to 12 with a step of 2.
It is observed that for any Card($\bm{B}_C$)$>0$, upgrading critical branches ranked by $K_i$ values can reduce CFR more greatly than by the structure vulnerabilities based indexes.
Though, upgrading critical branches identified by the structure vulnerabilities based indexes can effectively suppress CFR as well, they still seem less effective compared with the proposed $K_i$ value.
Thus, like the self-validation, this cross-validation also verifies the reasonability of the $K_i$ value to reflect branches' importance in cascading failure propagation.
Fig. \ref{fig:cross_validation_IEEE118}(b) shows the interval risk histograms for the case that Card($\bm{B}_C$)=12.
It is shown that CFR decreases from 29.81 MW to 6.93 MW by $K_i$ value based identification, to 15.36 MW by betweenness based identification, to 15.96 MW by electrical betweenness based identification and to 20.21 MW by extended betweenness based identification.
Further, we can observe that the risk of large blackouts (load loss$>30\%$) decreases significantly when all three structure vulnerabilities based identifications are utilized, which, however, have little effect on the risks of small blackouts (load loss$<10\%$) and medium blackouts ($10\%<$load loss$<30\%$).
\begin{figure}
\centering
\subfigure[]{\includegraphics[width=0.5\textwidth]{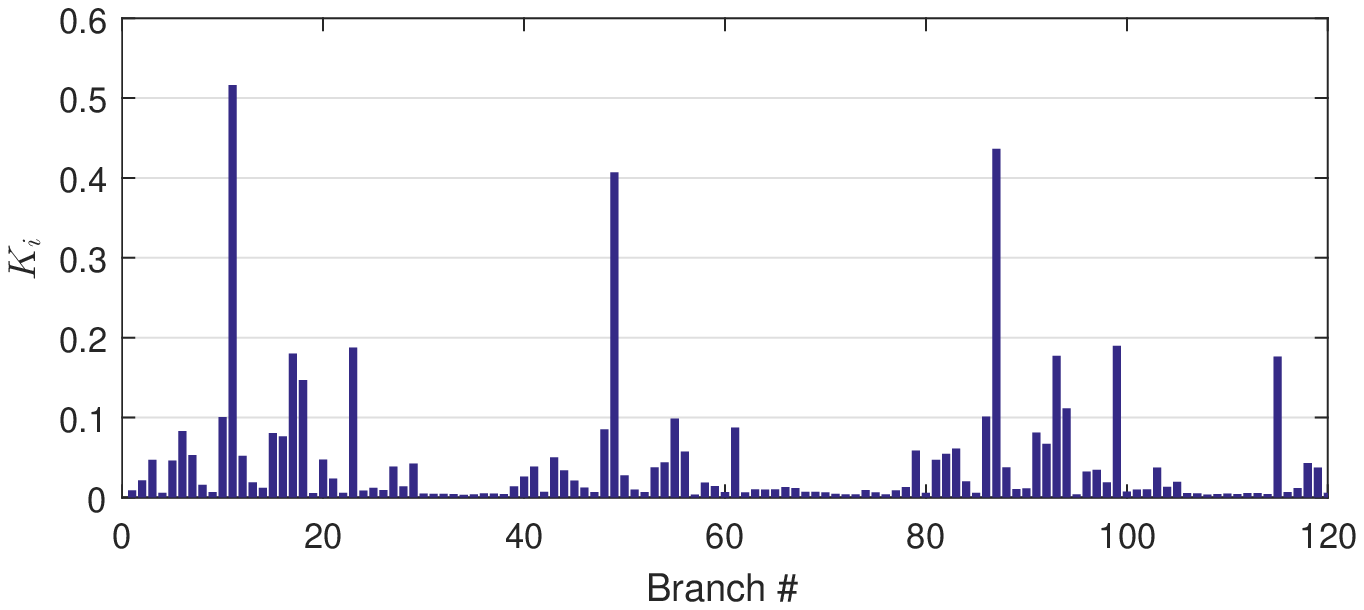}}
\subfigure[]{\includegraphics[width=0.5\textwidth]{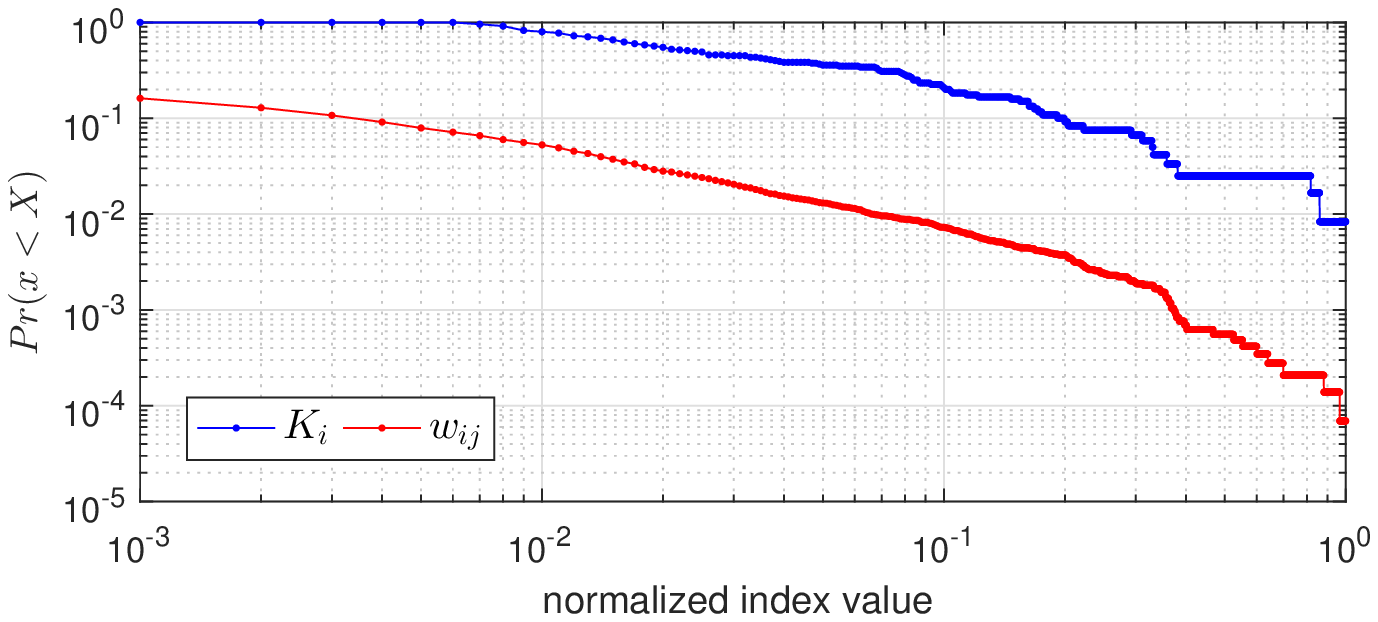}}
\caption{Index values of the identification results for IEEE RTS96 system, in which (a) shows the histogram of $K_i$ values across all branches marked from 1 to 120; (b) shows the distributions of $K_i$ and $w_{ij}$ which are normalized with min-max scaling.}\label{fig:index_distribution_RTS96}
\end{figure}
\subsection{Tests on IEEE three area RTS-96 system}
To further validate the proposed approach, case studies are also performed on the IEEE three area RTS-96 system, whose parameters can be found in \cite{RTS96_Test_System}.
It has 120 branches and 8550 MW load.
To increase the operation stress, each load is increased by $15\%$, and the value of $f_{lim1,l}$ is reduced to $70\%$ of the original transmission capacity limit for all branches.
Other parameters and pre-settings of simulation are the same with the cases on IEEE 118-bus system.
The histogram of $K_i$ values of all 120 branches are shown in Fig. \ref{fig:index_distribution_RTS96}(a).
It is observed that the largest $K_i$ value is bigger than $0.5$, and there are only 11 branches whose $K_i$ values exceed 0.3.
Apart from these high ranking branches, all others' $K_i$ values are below 0.1.
Distributions of normalized $K_i$ and $w_{ij}$ values are depicted in Fig. \ref{fig:index_distribution_RTS96}(b).
Similar to the situation of IEEE 118 bus system, both these two metrics of RTS79 system are also observed in power law distributions, which is another support to the awareness that only a small part of the branches have much higher importance than others.

Procedures of the validation for the identification results are the same with the cases on IEEE 118 bus system.
First, whole 120 branches of IEEE RTS96 system are ranked in a descending order in terms of their $K_i$ values.
Tab. \ref{tab:Branch_Ranking_HITS_RTS96} shows three typical groups of branches, i.e., \emph{top ranking branches}, ranking $1^{st}$ to $12^{th}$, \emph{middle ranking branches}, ranking $15^{th}$ to $26^{th}$, and \emph{bottom ranking branches}, ranking $109^{th}$ to $120^{th}$.
$K_i$ values of these three groups locate in the ranges $[0.1021,0.5171]$, $[0.0511, 0.0882]$ and $[3.62\times 10^{-3}, 4.46\times 10^{-3}]$ respectively.
Meanwhile, as comparison, top 12 branches of the rankings according to betweenness, electrical betweenness and extended betweenness are respectively shown in Tab. \ref{tab:Betweenness_Branch Ranking_RTS96}.
Self-validation and cross-validation are also made with the same procedures as in the cases on IEEE 118 bus system.
\begin{table}
\centering
\renewcommand{\arraystretch}{1.2}
\caption{Branch ranking with the proposed approach for IEEE RTS96 system}\label{tab:Branch_Ranking_HITS_RTS96}
\begin{tabular}{ccc|ccc}
\hline
\hline
Rank&Branch&$K_i$&Rank&Branch&$K_i$\\
\hline
1&11(107-108)&$0.5171$&21&83(303-324)&$0.0623$\\
2&87(307-308)&$0.4405$&22&82(303-309)&$0.0583$\\
3&49(207-208)&$0.4188$&23&12(108-109)&$0.0536$\\
4&23(114-116)&$0.1956$&24&56(211-213)&$0.0535$\\
5&99(314-316)&$0.1838$&25&7(103-124)&$0.0517$\\
6&115(107-203)&$0.1699$&26&79(301-305)&$0.0511$\\
7&17(110-112)&$0.1689$&...&...&...\\
8&93(310-312)&$0.1595$&109&70(218-221)&$4.46\times10^{-3}$\\
9&18(111-113)&$0.1501$&110&38(121-122)&$4.38\times10^{-3}$\\
10&94(311-313)&$0.1142$&111&71(218-221)&$4.27\times10^{-3}$\\
11&86(306-310)&$0.1046$&112&31(117-122)&$4.26\times10^{-3}$\\
12&10(106-110)&$0.1021$&113&32(118-121)&$4.07\times10^{-3}$\\
...&... &...&114&33 (118-121)&$3.93\times10^{-3}$\\
15&91(309-312)&$0.0882$&115&19(111-114)&$3.85\times10^{-3}$\\
16&6(103-109)&$0.0862$&116&109(318-321)&$3.71\times10^{-3}$\\
17&48(260-210)&$0.0836$&117&35(119-120)&$3.69\times10^{-3}$\\
18&61(214-216)&$0.0835$&118&111(319-320)&$3.68\times10^{-3}$\\
19&92(310-311)&$0.0791$&119&80(302-304)&$3.66\times10^{-3}$\\
20&16(110-111)&$0.0753$&120&73(219-220)&$3.62\times10^{-3}$\\
\hline
\hline
\end{tabular}
\vspace{-3mm}
\end{table}
\begin{table}
\centering
\renewcommand{\arraystretch}{1.2}
\caption{Branch ranking by structure vulnerability based metrics for IEEE RTS96 system}\label{tab:Betweenness_Branch Ranking_RTS96}
\begin{tabular}{cccccc}
\hline
\hline
\multicolumn{2}{c}{\emph{Betweenness}}&\multicolumn{2}{c}{\emph{Electrical Betweenness}}&\multicolumn{2}{c}{\emph{Extended Betweenness}}\\
\cmidrule(lr){1-2}\cmidrule(lr){3-4}\cmidrule(lr){5-6}
Rank&Branch&Rank&Branch&Rank&Branch\\
\hline
1&119(318-223)&1&119(318-223)&1&119(318-223)\\
2&120(323-325)&2&118(325-121)&2&118(325-121)\\
3&118(325-121)&3&120(323-325)&3&120(323-325)\\
4&104(316-317)&4&67(216-219)&4&67(216-219)\\
5&106(317-318)&5&29(116-119)&5&29(116-119)\\
6&24(115-116)&6&117(123-217)&6&105(316-319)\\
7&62(215-216)&7&62(215-216)&7&62(215-216)\\
8&67(216-219)&8&116(113-215)&8&104(316-317)\\
9&99(314-316)&9&104(316-317)&9&116(113-215)\\
10&95(311-314)&10&105(316-319)&10&66(216-217)\\
11&61(214-216)&11&66(216-217)&11&117(123-217)\\
12&29(116-119)&12&106(317-318)&12&28(116-117)\\
...&...&...&...&...&...\\
\hline
\hline
\end{tabular}
\vspace{-3mm}
\end{table}
\begin{figure}
\centering
\subfigure[]{\includegraphics[width=0.5\textwidth]{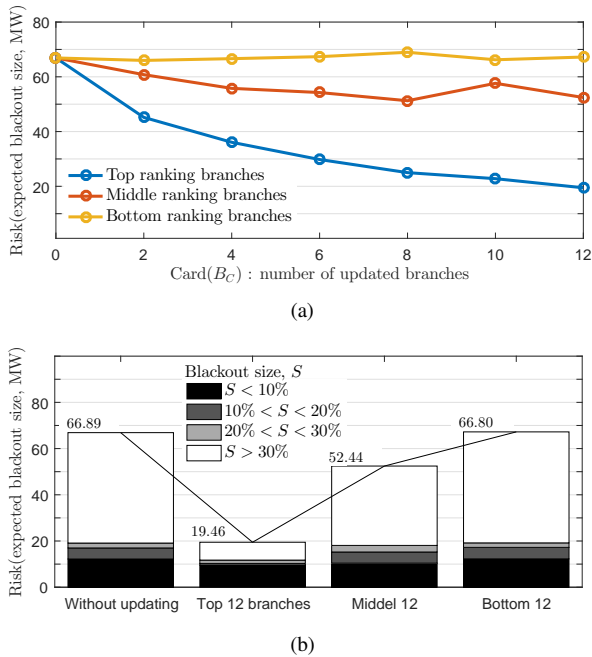}}
\subfigure[]{\includegraphics[width=0.5\textwidth]{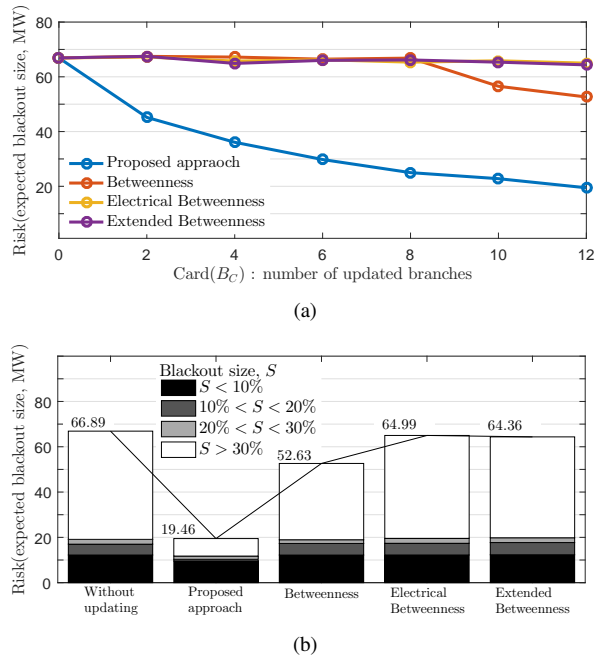}}
\caption{Results of self-validation on IEEE RTS96 system. (a) shows the performance comparisons of branch upgrading plans with top-ranking, middle-ranking and bottom-ranking branches identified by the proposed approach respectively; (b) shows the interval risk histograms of different upgrading plans when Card$(\bm{B}_C)=12$.}\label{fig:self_validation_RTS96}
\end{figure}

\begin{figure}
\centering
\subfigure[]{\includegraphics[width=0.5\textwidth]{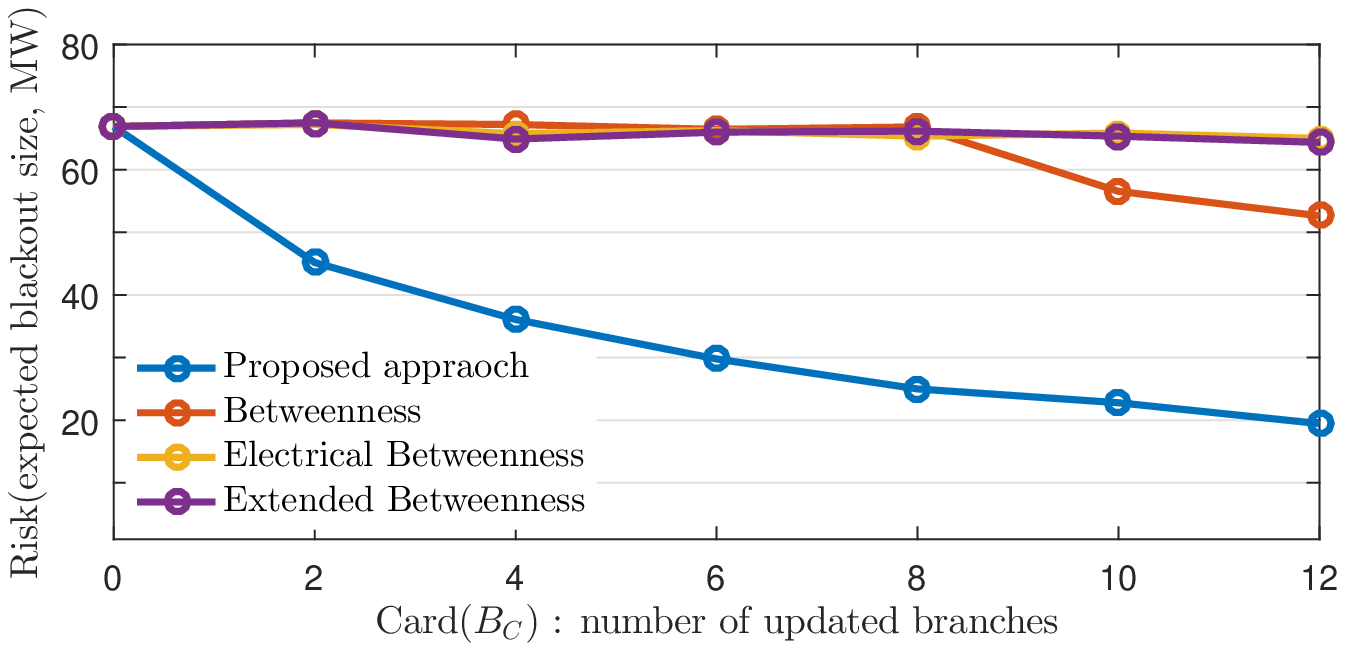}}
\subfigure[]{\includegraphics[width=0.5\textwidth]{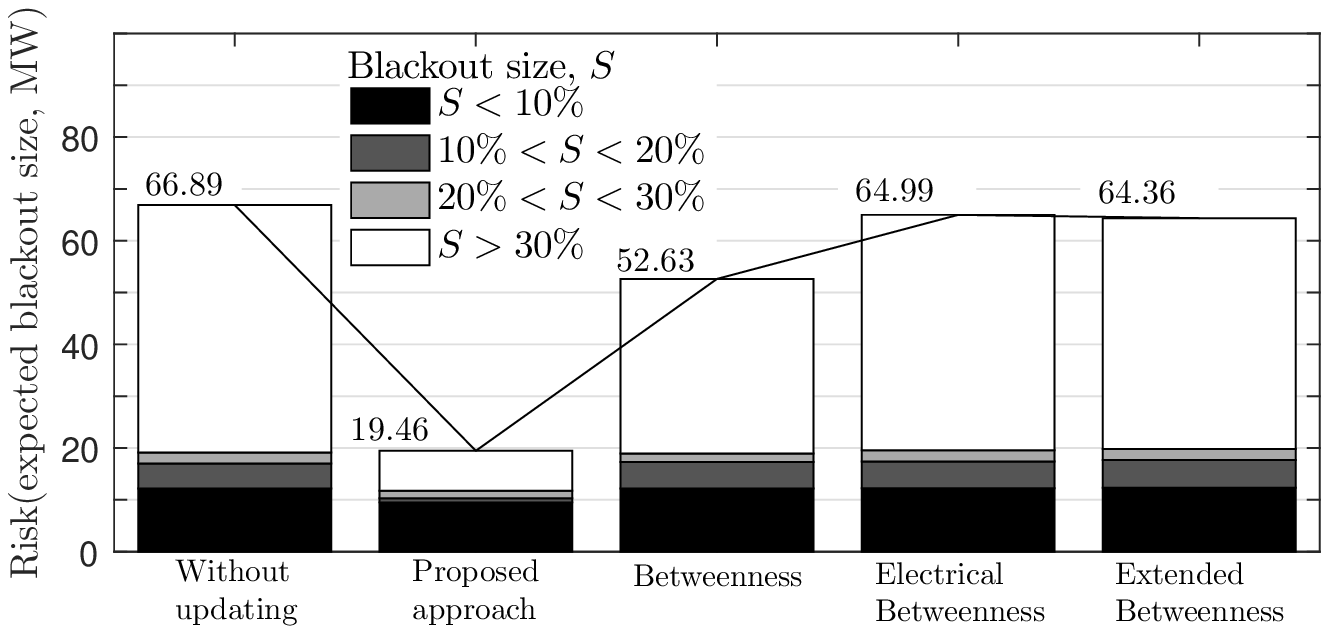}}
\caption{Results of self-validation on IEEE RTS96 system. (a) shows the performance comparisons of branch upgrading plans with top-ranking, middle-ranking and bottom-ranking branches identified by the proposed approach respectively; (b) shows the interval risk histograms of different upgrading plans when Card$(\bm{B}_C)=12$.}\label{fig:cross_validation_RTS96}
\end{figure}

Fig. \ref{fig:self_validation_RTS96} and Fig. \ref{fig:cross_validation_RTS96} demonstrate the results of self-validation and cross-validation respectively, from which several observations can be made.
First, as shown in Fig. \ref{fig:self_validation_RTS96}(a), with the increment of the number of upgraded branches (Card($\bm{B}_C$) increases from 0 to 12), CFR decreases monotonically with a relatively faster speed if the top ranking branches are selected, while when middle or bottom ranking branches are selected, CFR decreases much slower and even fluctuates when Card($\bm{B}_C)\geq 8$.
Thus, this observation validate that it is reasonable to find out the critical branches by the proposed $K_i$ value based identification.
Further, for a specific case that Card($\bm{B}_C)=12$, it is seen from Fig. \ref{fig:self_validation_RTS96}(b) that
CFR can be reduced from 66.89 MW (without upgrading) to 19.46 MW by upgrading top 12 branches, to 52.44 MW by upgrading middle 12 branches and to 66.80 MW by upgrading bottom 12 branches.
The risk of large blackouts (load loss$>30\%$) accounts for a large propagation in original CFR, and can be greatly reduced by upgrading top 12 branches, while upgrading other two branch groups seems much less effective.
Second, it is found from Fig. \ref{fig:cross_validation_RTS96}(b) that upgrading the ``critical branches'' identified by the three structure vulnerabilities based metrics shows little influence on CFR when Card($\bm{B}_C)$ increases.
Through exploring the interval risk histograms shown in Fig. \ref{fig:cross_validation_RTS96}(a), it is acquired that for the case that Card($\bm{B}_C)=12$, upgrading the ``critical branches'' identified by betweenness can reduce CFR to 52.63 MW, within which, however, the risk of large blackouts (load loss$>30\%$) does not decreases that much.

To summary, above self-validations and cross-validations on two typical benchmarks verify that it is reasonable to measure the branches' importance with the proposed $K_i$ value, and the identified critical branches are more preferred candidates in branch capacity upgrade to suppress CFR.

\section{Conclusion and Discussion}
This paper has proposed an effective way to identify the critical branches that have higher importance in cascading failure propagation.
The proposed analytic method for the statistics of cascading failure chains is a reliable tool to extract useful information from simulation results.
Detailed validations verify that the critical branches identified with the approach proposed in this paper are more favorable candidates for the capacity expansion problem.
Thus, our approach can effectively reduce the computational burden of the decision making on the capacity upgrading.
In the future, we will extend the research in the integration of the proposed approach with the optimal capacity expansion model.
There are several works along this direction could be extended, e.g., the cross validation of the proposed approach with the transient stability integrated cascading failure simulators, the application in large power grid with the consideration of slow evolution process, and the heuristic decision making with the combination of the proposed approach and the BCU model.
\section*{Acknowledgment}
The authors would like to thank the support by the National Natural Science Foundation (51190105, 51277135) and the State Grid Technology Project (XT71-12-011) of China.
\ifCLASSOPTIONcaptionsoff
  \newpage
\fi
\bibliographystyle{IEEEtran}
\bibliography{Reference}

\end{document}